\documentclass[prb,groupedaddress,preprint,showpacs,floatfix,12pt]{revtex4-1}
\usepackage{subfigure}
\usepackage{graphicx}
\usepackage{dcolumn}
\usepackage{bm}
\usepackage{amssymb}
\usepackage{amsmath}
\usepackage{epsfig}
\begin{document}
\title{Interpreting the 4-index Notation for Hexagonal Systems}
\date{\today}
\author{Philip B. Allen}
\affiliation{Department of Physics and Astronomy, Stony Brook University, Stony Brook, New York 11794-3800, USA}

\begin{abstract}
A four index notation ({\it e.g.} $(10\bar{1}1)$) is often used to denote reciprocal lattice vectors or crystal faces
of hexagonal crystals.  The purposes of this notation have never been fully explained.  This note clarifies the underlying
mathematics of a symmetric overcomplete basis.   This simplifies and improves the usefulness of the notation.
\end{abstract}

%
%
\maketitle

%
%
\section{Introduction}
A reciprocal lattice vector (RLV) has the form $\vec{G}=n_A\vec{A}+n_B\vec{B}+n_C\vec{C}$, where
$\vec{A},\vec{B},\vec{C}$ are primitive translations of the reciprocal lattice.
The integer ``indices'' $(n_A,n_B,n_C)$ were originally introduced as the Miller indices, which are reciprocals of the
intersection points (in units of $a=|\vec{a}|$, etc.), of a crystal lattice plane with the axes along the 
primitive translation vectors $\vec{a},\vec{b},\vec{c}$ of the direct lattice.
The ``4-index'' notation  $(10\bar{1}1)$ in a hexagonal crystal denotes an RLV with 
$(n_A,n_B,n_C)=(1,0,1)$.  The extra, or 
third, index $(\bar{1}=-1)$ in $(10\bar{1}1)$ is redundant.  This index must equal the 
negative sum of the first two indices.  The notation restores symmetry between equivalent directions 
which is lost if the third index is omitted.  This notation has long been used by crystallographers, dating back 
at least to Fedorov in 1890\cite{Fedorov1890}.  An extra Miller index is natural in hexagonal symmetry, because, 
rather than two $ab$-plane axes $\vec{a},\vec{b}$ at 120$^\circ$, the three symmetrical axes
$\vec{a}_1,\vec{a}_2,\vec{a}_3$, shown in Fig. 1, are natural.  The plane of atoms intersects all
three axes.  The reciprocal intersection points $(hki)$ are forced by trigonometry to obey the rule $h+k+i=0$.
The proof\cite{Donnay1947} is hinted in Fig. 1.

Even before Bragg scattering was observed and explained (1912-13),
the mathematical concept of a dual or reciprocal lattice was used\cite{Gibbs1901}.  After 1912, 
physicists recognized
the RLV as an x-ray momentum transfer.  The ``indices'' which label planes  seem
secondary unless we are specifically studying an atomic plane.  The redundant index may seem
only a nuisance.  Since the advantages of the 4-index notation are not always
understood, it is common to omit the third index, as would be done in crystal systems which lack 
$120^{\circ}$ rotations.  This note is written in the belief that, once the underlying idea is clearly understood,
the four index notation is natural.  It can be used to some advantage to label RLV's
$(hkil)$ and also to label directions $[stru]$ (called ``zones'' when the direction perceived as the common axis
of a family of planes) in the direct lattice.  However, the 4-index labeling of directions $[stru$ that
emerges in my analysis makes a subtle improvement over the one in use in electron microscopy\cite{Edington1976}.
Otte and Crocker\cite{Otte1965} discuss notation carefully, but with a different aim.

\begin{figure}[h]
\includegraphics[width=12cm]{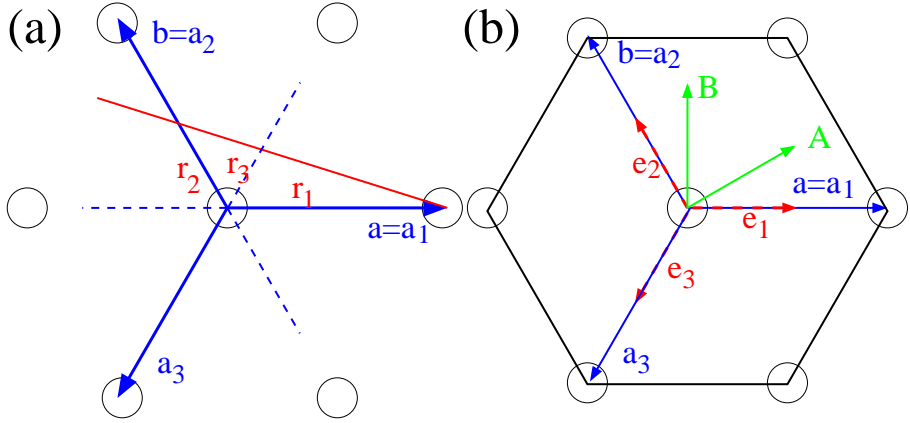}
\caption{(a) $ab$-plane of a hexagonal crystal, showing $a=a_1$ and $b=a_2$ axes, 
the third axis ($a_3$).  In red is shown the line of
intersection of a plane of atoms with the $ab$ plane.  
This red plane intersects the $a_1$ axis at $r_1$, the $a_2$ axis at $r_2$, and the $a_3$ axis
at $-r_3$.  It is always possible to choose the origin such that the intersection is at $r_1=1$.  Then the $120^\circ$ geometry guarantees that $1/r_3 = 1/r_2 + 1$.
(b) Plaquette of 2-d hexagonal crystal, showing primitive translations $\vec{a},\vec{b}$ of the
direct lattice and $\vec{A},\vec{B}$ of the reciprocal lattice.}
\label{fig:hex}
\end{figure}

Here is the main idea.  In an
hexagonal crystal, define four unit vectors, with the fourth ($\hat{e}_4$) pointing along the $c$-axis.  The other
three lie in the $ab$-plane, at $120^{\circ}$ to each other, as shown in Fig.\ref{fig:hex}.
It is conventional to have them point
in the directions of primitive translations of the lattice.  Then an arbitrary 3-vector $\vec{v}$ is written with four
components, as
\begin{equation}
\vec{v} \rightarrow |v\rangle =\left(\begin{array}{l}
\vec{v}\cdot\hat{e}_1 \\  \vec{v}\cdot\hat{e}_2 \\ 
\vec{v}\cdot\hat{e}_3 \\  \vec{v}\cdot\hat{e}_4
\end{array}\right) = \left( \begin{array}{l}
v_1 \\ v_2 \\ v_3 \\ v_4 \end{array} \right).
\label{eq:4v}
\end{equation} 
If $\vec{v}$ is a position vector $\vec{R}$ in ``real'' space, then it can be written as
\begin{equation}
\vec{R}  =\left(\begin{array}{l}
n_1 a \\  n_2 a \\  n_3 a \\  n_4 c
\end{array}\right) \rightarrow [n_1 n_2 n_3 n_4],
\label{eq:4R}
\end{equation} 
where the indices $n_i$ are dimensionless, and squared brackets conform to the crystallographic convention 
for directions in coordinate space.  
If $\vec{R}$ is a translation vector of the lattice, then the indices are integers.
If $\vec{v}$ is a vector of the reciprocal lattice $\vec{Q}$, 
then it can be written
\begin{equation}
\vec{Q}  =2\pi\left(\begin{array}{l}
n_1/ a \\  n_2/ a \\  n_3 /a \\  n_4 /c
\end{array}\right) \rightarrow (n_1 n_2 n_3 n_4),
\label{eq:4G}
\end{equation} 
where again the indices are dimensionless, and rounded brackets conform to convention.
If $\vec{Q}$ is a translation vector $\vec{G}$ of the reciprocal lattice, then the dimensionless indices are
integers, and the shorthand $\vec{G}=(n_1 n_2 n_3 n_4)$ is common.  In this paper, I will stick to a convention
that vectors written in column form as in Eqs.(\ref{eq:4R},\ref{eq:4G}), or in row form with components
separated by commas, contain the full dimensioned components, $\vec{v}\cdot\hat{e}_i$.  Dimensionless
abbreviations are never implied except in the index notations $[n_1 n_2 n_3 n_4]$ and $(n_1 n_2 n_3 n_4)$
without commas.  The term ``index'' will always refer to a dimensionless version of a component of a vector.

In Dirac notation, $|v\rangle$ indicates a column vector, 
while $\langle v|$ indicates a row vector.  However, in this paper notation switches
between two and three dimensions and between 2-vector,
 3-vector and 4-vector systems.  The Dirac notation is mostly avoided to reduce ambiguity.
When a vector is written alone, it can be assumed a column vector.  When 
written with another vector in a dot product $\vec{u}\cdot\vec{v}$, the left vector is a row
vector and the right vector a column vector.  When written as a dyad $\vec{v}\vec{v}$,
the left vector is  a column vector and the right vector a row vector.  When written
in a line of text with commas, the vector $(a,b,c,d)$ is probably a column vector written sideways
to save space.

The 4-vector ``basis'' is overcomplete.  It has both conventional and unconventional features.  
The basis vectors are, in 4-component notation
\begin{equation}
\hat{e}_1 \rightarrow
\frac{1}{2}\left(\begin{array}{r} 2 \\  -1 \\ -1 \\  0 \end{array}\right), \ \
\hat{e}_2 \rightarrow
\frac{1}{2}\left(\begin{array}{r} -1 \\  2 \\ -1 \\  0 \end{array}\right), \ \
\hat{e}_3 \rightarrow
\frac{1}{2}\left(\begin{array}{r} -1 \\  -1 \\ 2 \\  0 \end{array}\right), \ \
\hat{e}_4 \rightarrow \left(\begin{array}{l}
0 \\  0 \\ 0 \\  1 \end{array}\right).
\label{eq:4units}
\end{equation} 
Note that the sum of the first three indices is always zero, which follows from the
fact that the first three unit vectors add to zero, $\hat{e}_1 +\hat{e}_2 +\hat{e}_3 =0$.
It is perhaps disconcerting that the unit vectors $\hat{e}_1,\hat{e}_2,\hat{e}_3$ do
not ``look like'' unit vectors in the 4-vector notation, but this is a consequence of
overcompleteness.  For example, the second component of the 4-vector that
represents $\hat{e}_1$, is, according to Eq.(\ref{eq:4v}), $\hat{e}_1 \cdot \hat{e}_2 
=\cos 120^\circ= -1/2$.

The notation makes symmetry explicit. The six translations
have their first three indices chosen by the rule, 
organize three integers chosen from $\pm1$ and $\pm2$, in all ways such that they add to zero.
Because of the symmetrical mathematics, the scalar 
product ({\it e.g.} $\vec{u}\cdot\vec{v}=u_x v_x + u_y v_y + u_z v_z$)
in 4-vector form is simple but unconventional:
\begin{equation}
\vec{u}\cdot\vec{v}=\frac{2}{3}(u_1 v_1 + u_2 v_2 +u_3 v_3) + u_4 v_4.
\label{eq:dotprod}
\end{equation}
This can easily be verified, and will be explained in Sec. III.

\section{Conventional (``bi-orthogonal'') Basis}

When one drops the redundant third index from the 4-index notation, the remaining
indices express vectors in a non-orthogonal basis.
There is perfectly good mathematics behind this, but it hides symmetry and simplicity.  In a hexagonal
system, these disadvantages are removed by the 4-index system.  In systems of
lower symmetry, where translation vectors are not at $90^\circ$ and $120^\circ$
to each other, such simplification is not available.  In this section, the mathematics of non-orthogonal
basis sets is reviewed.

To simplify, the $c$-direction is now omitted.  The discussion thus
refers to crystallography of hexagonal crystals in two dimensions.  The third dimension
returns in Sec. IV.  In the 2-d space of the $ab$ plane, any two 
vectors that are not parallel or antiparallel can be chosen for a basis.  The most
obvious choices are the primitive translations $\vec{a}$ and $\vec{b}$, or alternately,
the primitive translations $\vec{A}$ and $\vec{B}$ of the reciprocal lattice,
defined as $\vec{A}=(2\pi/V)\vec{b}\times\vec{c}$, and similarly for $\vec{B}$.
$V$ is the volume of the three-dimensional unit cell, $V=(\sqrt{3}/2)a^2 c$.
Then we have the usual vector relations $\vec{A}\cdot\vec{a}=2\pi=\vec{B}\cdot\vec{b}$
and $\vec{A}\cdot\vec{b}=0=\vec{B}\cdot\vec{a}$.  These relations indicate that the
basis sets $(\vec{a},\vec{b})$ and $(\vec{A},\vec{B})$ are bi-othogonal.

It not always mentioned in texts that the primitive direct lattice vectors and the 
primitive RLV's are examples of the mathematical notion of
bi-orthogonal basis vectors.  The important property is completeness:  any vector can be expressed
as a unique linear combination $\vec{v}=v_a \vec{a} + v_b \vec{b}$.   Note that
$\vec{v}\cdot\vec{a}=v_a a^2+v_b \vec{a}\cdot\vec{b}$, 
where $\vec{a}\cdot\vec{b}$  equals $-a^2 /2$ in hexagonal systems.  The coefficients $v_a$
and $v_b$ are found by solving a $2\times 2$ linear system in $\vec{v}\cdot\vec{a}$ 
and $\vec{v}\cdot\vec{b}$.  A nice aspect of bi-orthogonality is that it 
diagonalizes the $2\times 2$ system and gives simple formulas 
for the coefficients, namely $v_a = \vec{A}\cdot\vec{v}/2\pi$, 
$v_b = \vec{B}\cdot\vec{v}/2\pi$.

It is arbitrary which basis (direct space or reciprocal) is taken to be
primary and which to be dual.  Thus, an arbitrary vector $\vec{v}$
has an alternate representation, $\vec{v}=v_{A} \vec{A} + v_{B} \vec{B}.$
The coefficients are
$v_{A} = \vec{a}\cdot\vec{v}/2\pi$, $v_{B} = \vec{b}\cdot\vec{v}/2\pi$.
The inner product $\vec{u}\cdot\vec{v}$ of two vectors is not given simply
by $v_a^2 + v_b^2$, but involves also the cross term $v_a v_b$.
The simple formula is $\vec{u}\cdot\vec{v}=2\pi(u_{A} v_a + u_{B} v_b)$,
where the row vector is expressed in the basis dual to the one chosen 
for the column vector.
An equivalent formula is $\vec{u}\cdot\vec{v}=2\pi(u_a v_{A} + u_b v_{B})$. 
A compact mathematical representation of these relations is given by the equation
$|A\rangle\langle a| + |B\rangle\langle b| = 2\pi{\bf {\mathsf 1}}$, or by the alternate
equation $|a\rangle\langle A| + |b\rangle\langle B|= 2\pi{\bf {\mathsf 1}}$.  Here the
notation $|v\rangle$ means a 2-vector, and ${\bf {\mathsf 1}}$ is the 
$2\times 2$ unit matrix.  If written as a 2-component column vector,
the basis should either be orthonormal, or if non-orthonormal, one has to be careful
to use the direct and dual basis for the column vector $|v\rangle$ 
and the row vector $\langle v|$.  In dyad notation, the relations are
\begin{equation}
\vec{A}\vec{a} + \vec{B}\vec{b} =\vec{a}\vec{A} + \vec{b}\vec{B}= 2\pi{\bf {\mathsf 1}}.
\label{eq:unityAa}
\end{equation}
This formula is called ``the completeness
relation'' or, equivalently, ``the decomposition of unity.''  Although this gives elegant
formulas for inner products in non-orthogonal basis sets, these formulas are not likely to be used
unless the vectors $\vec{u}$ and $\vec{v}$ belong separately, one to direct, and the other to reciprocal
space.  Then the formulas are obvious.  One has no trouble realizing that $\vec{Q}\cdot\vec{r}
=(Q_{A} \vec{A} + Q_{B} \vec{B})\cdot(r_a \vec{a} + r_b \vec{b})$ is equal to 
$2\pi(Q_{A} r_a +Q_{B} r_b)$.

\section{Overcomplete symmetrical basis}

For two dimensional vectors, or the $ab$-plane components of 3-d vectors of an hexagonal crystal,
the overcomplete symmetrical basis is the three vectors $\hat{e}_1,\hat{e}_2, \hat{e}_3$
which lie at $120^\circ$ to each other, as shown in Fig.\ref{fig:hex}.  
The key relationship is the dyad formula
\begin{equation}
\frac{2}{3}(\hat{e}_1\hat{e}_1 + \hat{e}_2\hat{e}_2 + \hat{e}_3\hat{e}_3)={\bf \mathsf{1}}.
\label{eq:unity3e}
\end{equation}
This decomposition of unity is a very nice alternative to Eq.(\ref{eq:unityAa}).
It is no longer necessary to have dual sets of direct and reciprocal lattice vectors.
The three vectors $\hat{e}_i$ are self-dual.  The scalar product of any two 2-vectors
can be written as 
\begin{equation}
\vec{u}\cdot\vec{v} = \vec{u}\cdot\mathsf{1}\cdot\vec{v}=\frac{2}{3}(\vec{u}\cdot\hat{e}_1
\hat{e}_1\cdot\vec{v} + \vec{u}\cdot\hat{e}_2\hat{e}_2\cdot\vec{v}
+\vec{u}\cdot\hat{e}_3\hat{e}_3\cdot\vec{v}).
\label{eq:3dotprod}
\end{equation}
This formula can be written in an overcomplete 3-vector notation as
\begin{equation}
\vec{u}\cdot\vec{v} = \frac{2}{3}(u_1 v_1 + u_2 v_2 +u_3 v_3)=
\frac{2}{3}\left(\begin{array}{lll}u_1 & u_2 & u_3\end{array}\right)
\left(\begin{array}{r}v_1 \\  v_2 \\ v_3\end{array}\right),
\label{eq:2dotprod}
\end{equation}
where the usual array multiplication rule is obeyed.  Unlike the case of the
biorthogonal basis, here there is no need for care about whether $\vec{u}$
or $\vec{v}$ is in direct or reciprocal space, or whether the primary or the dual
basis is implied.  The formula works for any two vectors.  

It is worth emphasizing one subtlety.  In the bi-orthogonal basis, when a column
vector has indices $[v_a v_b]$, this means that $\vec{v}=v_a \vec{a} + v_b \vec{b}$,
even though $v_a \ne \vec{v}\cdot\vec{a}$.  Similarly, if the indices are written $(v_A v_B)$,
this indicates a vector $\vec{v}=v_A \vec{A} + v_B \vec{B}$.
Very different relations hold true in the symmetric
overcomplete basis.  If the vector has components $v_1,v_2,v_3$, this means
that $v_1 = \vec{v}\cdot\hat{e}_1$, {\it etc.}, even though $\vec{v} \ne 
v_1 \hat{e}_1 +v_2 \hat{e}_2 + v_3 \hat{e}_3$.  The actual answer is
$\vec{v} = (2/3)(v_1 \hat{e}_1 +v_2 \hat{e}_2 + v_3 \hat{e}_3)$. 
This follows immediately from Eq.(\ref{eq:unity3e}).  It does not matter whether the
vector $\vec{v}$ is in direct space and indexed as $[v_1 v_2 v_3]$
or in reciprocal space and indexed as $(v_1 v_2 v_3)$.

Now let us express the reciprocal lattice vectors in the overcomplete symmetric
notation.  First, recall that since $\vec{A}\cdot\vec{a}=(\sqrt{3}/2)|\vec{A}|a=2\pi$,
we have $|\vec{A}|=(2/\sqrt{3})(2\pi/a)$.  Then we find $\vec{A}\cdot\hat{e}_1
=(\sqrt{3}/2)|\vec{A}|=2\pi/a$, $\vec{A}\cdot\hat{e}_2=0$, and $\vec{A}\cdot\hat{e}_3=-2\pi/a$.
Therefore we have
\begin{equation}
\vec{A}_1=\vec{A}=\frac{2\pi}{a}\left(\begin{array}{r}1\\0\\-1\end{array}\right), \ \ 
\vec{A}_2=\vec{B}=\frac{2\pi}{a}\left(\begin{array}{r}0\\1\\-1\end{array}\right), \ \ 
\vec{A}_3=\vec{B}-\vec{A}=\frac{2\pi}{a}\left(\begin{array}{r}-1\\1\\0\end{array}\right).
\label{eq:G3}
\end{equation}
The set of six symmetry-related smallest RLV's $(hki)$ are simply all permutations of
the three indices $1,0,-1$.  The general reciprocal lattice vector is written in various ways, as
\begin{equation}
\vec{G}=n_A\vec{A}+n_B\vec{B}\rightarrow (n_A n_B),
\label{eq:G2}
\end{equation}
\begin{equation}
\vec{G}\rightarrow\frac{2\pi}{a}\left[n_A\left(\begin{array}{r}1\\0\\ \bar{1}\end{array}\right)
+n_B \left(\begin{array}{r}0\\1\\ \bar{1}\end{array}\right)    \right] 
=\frac{2\pi}{a}\left(\begin{array}{c}n_A\\n_B\\-n_A-n_B\end{array}\right) \rightarrow (n_A n_B \  \overline{n_A+n_B}).
\label{eq:G33}
\end{equation}
It is amazing how similar the bi-orthogonal representation (Eq.\ref{eq:G2})
is to the overcomplete symmetric representation (Eq.\ref{eq:G33}) for RLV's.  
They involve different coordinate systems and rules, yet the former derives from
the latter by just dropping the extra third index, and the latter from
the former by adding an index which is the negative sum of the first two indices.

The additional advantages of the extra index representation are that it is completely
explicit (if $2\pi/a$ is included, the vector is completely specified), and there is no
ambiguity about the direct versus dual parts of the bi-orthogonal representation.

\section{Conclusions}

It is best to think of Eq.(\ref{eq:4v}) as a way of representing
the vector $\vec{v}$ but not to think of the components of the 4-vector as if they had
meaning similar to the ordinary 3-component notation.  In ordinary vector
notation, if the vector $\vec{v}$ is denoted $[v_a v_b v_c]$, 
then it has the formula $v_a\vec{a} + v_b \vec{b} + v_c \vec{c}$.  Or,
if it is denoted by 
$(v_A v_B v_C)$, 
then it has the formula $v_A\vec{A} + v_B \vec{B} + v_C \vec{C}$. 
In 4-index notation, vectors indexed as $(n_1 n_2 n_3 n_4)$
or  $[n_1 n_2 n_3 n_4]$ have the formulas
\begin{equation}
\vec{u} = (n_1 n_2 n_3 n_4)
=2\pi \left(\begin{array}{r} n_1/a \\ n_2/a \\ n_3/a \\ n_4/c\end{array}\right)
\rightarrow 2\pi\left(\frac{2}{3a}[n_1\hat{e}_1 + n_2 \hat{e}_2 + n_3\hat{e}_3] + \frac{1}{c}n_4 \hat{e}_4 \right).
\label{eq:vyes}
\end{equation}
\begin{equation}
\vec{v} = [n_1 n_2 n_3 n_4]
= \left(\begin{array}{r} n_1 a \\ n_2 a \\ n_3 a \\ n_4 c\end{array}\right)
\rightarrow \left(\frac{2a}{3}[n_1\hat{e}_1 + n_2 \hat{e}_2 + n_3\hat{e}_3] + c n_4 \hat{e}_4 \right).
\label{eq:vyes2}
\end{equation}

Note the unconventional factor 2/3.  
When written as an additive sum of primitive vectors proportional to $\hat{e}_i$, 
an arbitrary additive constant can be added to $u_1, u_2, u_3$ or
 $v_1,v_2,v_3$ with no algebraic or notational error.  For example,
\begin{equation}
\vec{v} =\frac{2a}{3}[(v_1+C)\hat{e}_1 +( v_2+C )\hat{e}_2 + (v_3+C)\hat{e}_3] + v_4 c \hat{e}_4
\ne \left(\begin{array}{c} (v_1+C)a  \\ (v_2+C)a \\ (v_3+C)a \\ v_4 c\end{array}\right)
\label{eq:vno}
\end{equation}
When written in overcomplete 4-vector notation,
$v_1, v_2, v_3$ are fixed numbers, necessarily adding to zero, and therefore
with no arbitrariness.  The constant $C$ cannot be added, even though the vector relation
$C(\hat{e}_1+\hat{e}_2+\hat{e}_3)=0$ is true.

The inner product of two vectors, in 4-vector notation, is
\begin{equation}
\vec{u}\cdot\vec{v}=\left(\begin{array}{llll}u_1, & u_2, & u_3,&u_4\end{array}\right)
\left(\begin{array}{rrrr} 2/3 &0 &0 &0\\ 0&2/3&0&0\\
0&0&2/3&0\\0&0&0&1\end{array}\right)
\left(\begin{array}{r}v_1 \\  v_2 \\ v_3\\v_4\end{array}\right),
\label{eq:3dotprod}
\end{equation}
 This is just an alternative way of writing Eq.(\ref{eq:dotprod}), that emphasizes
 the unconventional metric.  A factor
 $2/3$ occurs in the first three diagonal entries.  The metric is positive,
 so the inner product is safely defined.  
 Eq.(\ref{eq:3dotprod}) holds for all vectors provided the indices $u_i$ and $v_i$ are written in full rather 
 than abbreviated index form.  For the special case where one of $\vec{u},\vec{v}$ is in direct and the other
 in reciprocal space, the inner product, {\it modulo} $2\pi$, also has this form, {\it i.e.}
 $(hkil)\cdot[stru]=2\pi[(2/3)(hs+kt+ir)+lu]$, in the index form.  If both are direct or reciprocal space
 vectors, an extra factor $(c/a)^2$ or $(a/c)^2$ appears in the fourth term.
  
 The unconventional metric with the factor $2/3$ does not appear in previous literature.  
 If the 4-index notation is confined to indexes of planes and reciprocal lattice vectors, then the 
 current definition Eq.(\ref{eq:4v}) agrees with the one in the literature.  
 It is less common for a 4-index notation to be used for coordinate space
 directions.  Inner products are seldom discussed, and the factor $2/3$ doesn't arise.

For real space directions directions, the current definition Eq.(\ref{eq:4v})
differs from used in the literature\cite{Edington1976,Otte1965}.
Consider for example the lattice point $\vec{R}=n_a\vec{a}+n_b\vec{b}+n_c\vec{c}$.
Using Eqs.(\ref{eq:4v},\ref{eq:4R}), this is indexed as $[(n_a+\bar{n}_b/2) \ (n_b + \bar{n}_a/2) \ 
(\overline{n_a + n_b}/2) \ n_c]$.  The literature definition is different.  It requires 
first writing $\vec{R}=n_a\vec{a}_1+n_b\vec{a}_2+n_c\vec{c} -(1/3)(n_a+n_b)
(\vec{a}_1 + \vec{a}_2 + \vec{a}_3))$.   The extra term is zero but is added to make
the sums of the coefficients of $\vec{a}_1$, $\vec{a}_2$, and $\vec{a}_3$ add to zero.
By the literature definition, this vector is indexed as $[stru]$ where the 
indices are the coefficients of $\vec{a_1}$, etc., and $s+t+r=0$.  Thus we get $\vec{R} \rightarrow
[2(n_a+\bar{n}_b)/3 \ (2n_b + \bar{n}_a)/3 \ (\overline{n_a + n_b}/3) \ n_c]$.
The first three indices have been reduced by $2/3$.  
 It never seems to be mentioned that the definition was changed between direct and reciprocal space.
 The factor $2/3$ is now incorporated into the definition of the
 first three indices $str$ of $[stru]$ of direct space vectors,
 but not into the first three indices $(khil)$ of reciprocal space vectors.  Then the inner product of
 a direct space vector $[stru]$ and a reciprocal space vector $(hkil)$ is $2\pi(sh+tk+ri+ul)$, 
 without the extra $2/3$.
 The penalty is that inner products of two direct space vectors or two reciprocal space vectors must be
 computed with awkwardly different rules, whereas the unified definition offered here gives a simple unified
 (but unconventional) rule.  In order to retain the usual definitions, one could accept 
 as a compromise the {\it ad hoc} rule that
 whenever dimensionless real space indices $[stru]$ are written, they incorporate an extra factor $2/3$ beyond Eq.(\ref{eq:4v}),
 in the first three indices.  However, when written out in full column vector notation,
 there is no need for index notation, and such a compromise would be unwise.  Eq.(\ref{eq:4v})
 should be used, and the inner product rule Eq.(\ref{eq:dotprod}) applies.
 
 Overcomplete basis sets are not abnormal in physics.  They are used frequently for infinite-dimensional
 problems.  A quantum harmonic oscillator, for example, has an infinite complete orthonormal basis of eigenstates 
 $|n\rangle$, but the ``coherent state'' representation, which is overcomplete, is often preferable\cite{Negele1988}.  
 In finite-dimensional vector
 problems, hexagonal crystallography is not unique; symmetry and orthonormality may compete and
 suggest an overcomplete symmetric representation.  An example is electronic $d$-states, where cubic symmetry
 lifts 5-fold degeneracy into a 3-fold degenerate T$_{2g}$ manifold (spanned by orthonormal
 functions $yz$, $zx$, and $xy$) and the
 2-fold degenerate E$_g$ manifold (spanned by orthogonal functions $x^2-y^2$ and $3z^2 -r^2$.)  The T$_{2g}$ orthogonal
 basis is nicely adapted to the 4-fold rotations of cubic symmetry, whereas these same rotations mix the orthogonal E$_g$
 functions in an ugly way.  A cure\cite{Perebeinos2000} 
 is an overcomplete non-orthogonal basis such as $3x^2-r^2$, $3y^2 -r^2$, and $3z^2-r^2$.
 The mathematics of this representation is exactly the same as the symmetric overcomplete representation
 described here for hexagonal crystals.

\begin{acknowledgements}
 I thank my collaborators in the Solar Water Splitting Simulation Team (SWaSSiT)
 who helped me understand the $(10\bar{1}1)$ surface of wurtzite materials.
 M. Blume and A. G. Abanov made valuable comments.  I thank the CFN of
 Brookhaven National Laboratory for hospitality.  This work was supported
 by DOE grant DEFG0208ER46550.
 \end{acknowledgements}

\end{document}